\documentclass[floatfix,superscriptaddress,aps,prd,nofootinbib,preprint]{revtex4-1}
\usepackage{graphicx}
\usepackage{amsmath, mathtools}
\usepackage{amssymb}
\usepackage{bm}
\usepackage[normalem]{ulem}
\usepackage[colorlinks=true,citecolor=blue]{hyperref}
\usepackage{color}
\usepackage{xcolor}

\newcommand{\Tr}{{\rm Tr}}

\newcommand{\vN}{\text{vN}}
\newcommand{\union}{\cup}

\begin{document}
\count\footins=1000

\title{Energy dynamics, information
and heat flow in quenched cooling and the crossover from quantum to classical thermodynamics}

\begin{abstract}

The dynamics when a hot many-body quantum system is brought into instantaneous contact with a cold many-body quantum system can be understood as a combination of early time quantum correlation (von Neumann entropy) gain and late time energy relaxation.
We show that at the shortest 
timescales 
there is an energy 
increase in each system linked to the entropy gain, even though {equilibrium} thermodynamics does not apply. 
This energy increase is of quantum origin and results from the collective binding energy between the two systems.
Counter-intuitively, this implies that {also} the hotter of the two systems generically experiences an initial energy increase when brought into contact with the other colder system. 
In the limit where the  energy relaxation overwhelms the (quantum) correlation build-up, classical energy dynamics emerges 
where the energy in the hot system decreases immediately upon contact with a cooler system. 
We use both strongly correlated SYK systems and weakly correlated mixed field Ising chains to exhibit these characteristics, and comment on its implications for both black hole evaporation and quantum thermodynamics.

\end{abstract}

\author{V. Ohanesjan}
\email{ohanesjan@lorentz.leidenuniv.nl}
\affiliation{Instituut-Lorentz, $\Delta$-ITP, Universiteit Leiden, P.O. Box 9506, 2300 RA Leiden, The Netherlands}

\author{Y. Cheipesh}
\affiliation{Instituut-Lorentz, Universiteit Leiden}

\author{N. V. Gnezdilov}
\affiliation{Department of Physics, University of Florida, Gainesville, FL 32611, USA}

\author{A. I. Pavlov}
\affiliation{The Abdus Salam International Centre for Theoretical Physics (ICTP) Strada Costiera 11, I-34151 Trieste, Italy}

\author{K. Schalm}
\affiliation{Instituut-Lorentz, $\Delta$-ITP, Universiteit Leiden, P.O. Box 9506, 2300 RA Leiden, The Netherlands}


\maketitle

\section{Introduction}

The notion of entropy is more involved in quantum systems than in classical systems as it also 
includes the information of potential entanglement with another set of dynamical degrees of freedom. This can be another system with which it is (weakly) coupled,  the environment, or the measurement apparatus.
In classical equilibrium thermodynamics, the change in the entropy is associated with  
heat flow according to the Second Law while 
quantum mechanically the entropy can be changed by the quantum correlations in the system that may or may not necessarily involve heat flow. The field of quantum thermodynamics specifically pursues this question how 
work, heat and entropy 
are affected by quantum 
correlations including entanglement;
see e.g. \cite{Goold_2016,deffner2019quantum} for recent reviews. This 
field
is growing rapidly, even though these many-body entanglement effects are still less well understood than entanglement and decoherence in 
few-qubit systems.

Here we take a quantum thermodynamics point of view on non-equilibrium dynamics in many-body systems with two theoretical models as example: the Sachdev-Ye-Kitaev (SYK) model and a mixed field Ising chain. The Sachdev-Ye-Kitaev model has a computable non-Fermi liquid ground state
that is long-range many body entangled 
\cite{Sachdev1993Gapless,Kitaev2015Simple}. 
Through the holographic duality between anti-de-Sitter quantum gravity and matrix large $N$ quantum systems, such SYK models at finite temperature are also dual descriptions of black holes in anti-de-Sitter gravity 
\cite{Sachdev2015Bekenstein}. 
Using this duality
to study the profound question of black hole evaporation through Hawking radiation and its information flow \cite{Penington:2019npb,Almheiri2019Entropy,Penington:2019kki,Almheiri2019Page,chen2020evaporating}, recent studies have considered the quenched cooling of a hot thermal SYK state (the black hole) suddenly being able to ``evaporate'' into a cooler or even $T=0$ SYK state (the container for the e\-va\-po\-rated radiation)  \cite{Zhang2019Evaporation,Almheiri2019Universal,maldacena2020syk}.\footnote{Early work on SYK quenches is \cite{Eberlein:2017wah}. For other aspects of SYK dynamics, see this and citations thereof.} A surprising finding from the perspective of classical thermodynamics has been that these observe an initial energy {\em increase} \cite{Almheiri2019Universal, Zhang2019Evaporation,maldacena2020syk,larzul2022fast} in the hot subsystem, confirming results from preceding black hole evaporation studies \cite{Almheiri:2018xdw}.   
It was 
argued, using Schwinger-Keldysh field theory, that many relativistic continuum field theories will exhibit such an energy increase in the hot system when quench coupling two thermal states \cite{Almheiri:2018xdw,Almheiri2019Universal} even though a
fundamental proof or understanding  
was 
missing.
In particular, a quenched cooling 
between two two-level systems provides a counterexample \cite{Almheiri2019Universal}.\footnote{{The thermal state  of a two-level system is defined through its density matrix $\rho =  \frac{1}{Z} \sum_{n} |n\rangle e^{-\beta E_{n}}\langle n| $ with $n=\downarrow,\uparrow$ and $Z$ the appropriate normalization such that Tr$\rho=1$.}}  

In a recent article, we showed that quantum thermodynamics \cite{Goold_2016,deffner2019quantum} provides the universal explanation for this counterintuitive 
rise \cite{gnezdilov2021information}. In a quenched cooling protocol where a (hot) thermal quantum system with Hamiltonian $H_A$ is brought into instantaneous contact with a (cooler) thermal reservoir at $t=0$ through $H_{\text{total}}=H_A+H_B+\theta(t)H_{\text{int}}$, the change in the energy of the hot subsystem $A$ equals 
\begin{align}
\label{eq:1a}
    \Delta E_{A}(t) = T_{A}\Delta S_{\vN,A}(t) + T_AD(\rho_A(t)||\rho_{T_A})~.
\end{align}
Here $S_{\vN}=-\Tr(\rho_A\ln\rho_A)$ is the von-Neumann entropy of the reduced density matrix of the subsystem $A$: $\rho_A=\Tr_B\rho$; the energy of the subsystem $E_A(t)$ is the expectation value of its subsystem Hamiltonian $E_A =\Tr H_A\rho(t) = \Tr H_A\rho_A(t)$; and {$D(\rho_A(t)||\rho_{T_A}) = \Tr \rho_A(t) \log \left(\rho_A(t)/\rho_{T_A}\right)$ }is the relative entropy between the reduced density matrix of system $A$ and the initial thermal density matrix of $A$ at $t=0$. The change $\Delta E(t)=E(t)-E(0)$ is with respect to the same quantity at $t=0$. By symmetry an analogous relation holds for subsystem $B$.

As the relative entropy $D(\rho_A(t)||\rho_{T_A}) \geq 0$ is positive semi-definite, one arrives at an inequality 
that holds universally for any model Hamiltonian when such a quenched cooling protocol is considered 
\begin{align}
\label{eq:1}
    \Delta E_A(t) \geq T_A \Delta S_{\vN,A}(t)~.
\end{align}
In a quantum system the von-Neumann entropy can have a significant contribution from quantum correlations including
entanglement over and above the classical thermal 
entropy.
As the 
quantum correlations
between the system and the reservoir can only increase after a quench,
the quantum thermodynamic inequality Eq.\eqref{eq:1} can therefore force an associated {\em increase} in energy in system $A$ even if its initial energy density was higher. Moreover, in perturbation theory to leading order 
the inequality saturates as the contribution of the relative entropy is subleading and one can use the equality as a way to measure the von Neumann entropy in a quenched cooling protocol through the energy difference \cite{gnezdilov2021information}.

A common view on non-equilibrium phenomena is that at the shortest time scales the system is extremely sensitive to microscopic information, details of the quench protocol etc, and it is only the longest-time-scale-relaxation to equilibrium that is universal. Eq.\eqref{eq:1} surprisingly shows 
that it need not be so:
at the shortest possible non-equilibrium time scale there is still a notion of the first law that entropy is linked to energy, even though the standard first law in the absence of work $dE=TdS$ is relating  state functions regarding equilibria.

This positive contribution due to quantum correlations to the von Neumann entropy is present in {\em any}  quantum system, but our classical experience is that the energy in the hot system decreases directly upon contact because heat  must flow from hot to cold. What must happen to restore this intuition that the energy in the hot system decreases instantaneously is that the positive quantum correlation- and entanglement- 
contribution can be overwhelmed by the semi-classical heat and information flow from hot to cold. 
By studying quenched cooling in SYK models, where entanglement is very strong, and one-dimensional mixed field Ising chains, where entanglement can be made very weak, we exhibit this. Classical experience is restored in a particle-like system at high temperatures where entanglement is weak.

\section{Energy dynamics in quenched cooling}

The setup we study consists of two initially independent quantum subsystems $A$ and $B$ with Hamiltonians $H_A$ and $H_B$ respectively. Initially $(t<0)$, each subsystem is prepared in a thermal state at temperature $T_A$ and $T_B$, 
so the full system is in an uncorrelated product state:
\begin{gather}
    \rho_0=\rho_{T_A}\otimes \rho_{T_B} \nonumber
    \\
    \rho_{T_\alpha} = \frac{1}{Z_\alpha}e^{-H_\alpha / T_\alpha}, \quad \alpha =A,B.
\end{gather}
We will study the behaviour of the subsystems when they are brought into instantaneous contact at $t=0$ through an interaction Hamiltonian $H_{int}$. The complete setup is a closed system that evolves with the full Hamiltonian:
\begin{align}
    H_{\text{total}}=H_A+H_B+\theta(t)H_{\text{int}}~.
\end{align}
Motivated by current results presented in the introduction, we focus our interest on two different models:
\begin{itemize}
    \item Finite $N$ Majorana SYK with each subsystem governed by the Hamiltonian
    \begin{gather} \label{eq:MSYK}
        H_{\alpha} = i^{q/2} \sum_{j_1 \dots j_q=1}^{N_\alpha} J^{\alpha}_{j_1 \dots j_q} \psi_{j_{1}}^{\alpha} \dots \psi_{j_q}^{\alpha}  ~~~~~~~~\alpha=A,B
    \end{gather}
    where $q$ is same for both dots and can be either $q=2$ or $q=4$, further labeled as SYK$_2$ and SYK$_4$ respectively. The couplings are drawn from a Gaussian distribution with the following parameters:
    \begin{gather} \label{eq:SYK_Jcouplings}
        \langle J^{\alpha}_{j_1 \dots j_q }  \rangle =0,
        \quad
        \langle J^{\alpha}_{j_1 \dots j_q } J^{\alpha}_{j_1 \dots j_q }  \rangle = \frac{(q-1)! J ^2 }{N_\alpha^{q-1}}.
    \end{gather}
    Those two SYK dots are coupled through a two Majorana tunneling interaction which couplings are also sampled from a Gaussian distribution:\footnote{{We have taken a variance in $\lambda$ that is asymmetric in $N_A$ and $N_B$ to readily compare with \cite{Zhang2019Evaporation,Almheiri2019Universal}. These authors chose this such that the interaction stays relevant in the large $N_A$ limit.}}
    \begin{gather} \label{eq:MSYK_int}
    H_{int} = i\sum_{ij} \lambda_{ij} \psi_i^{A} \psi_j^{B},
    \\
    \langle \lambda_{ij} \rangle =0,
    \quad 
    \langle \lambda_{ij}^2 \rangle = \frac{\lambda^2}{N_B}.
    \end{gather}
    This system is analyzed with exact diagonalization and 
    averaged over
    $R={100}$ different coupling realizations. To reduce the number of free parameters we take two equal size dots $N_A=N_B \equiv N$.

    \item The 1D mixed field Ising model, also analyzed using exact diagonalization, with a particle-like contact interaction:
    \begin{gather} \label{eq:H_Ising}
        H_\alpha = -\sum_{i}^{ N_\alpha} \left(J Z_i^{\alpha} Z_{i+1}^\alpha + g X_i^\alpha + h Z_i^\alpha \right), ~~~~~~~~\alpha=A,B
        \\
        H_{\text{int}}^{(tunn.)} = -
        \lambda  (X+iY)_{N_A}^{A} (X-iY)_{1}^B +{h.c.} \label{eq:H_Ising_Interaction}
    \end{gather}

\end{itemize}
{Dimensionful parameters are expressed in $J$, which is usually set to $J=1$.}

\begin{figure}[!t]
\includegraphics[width=0.5\textwidth]{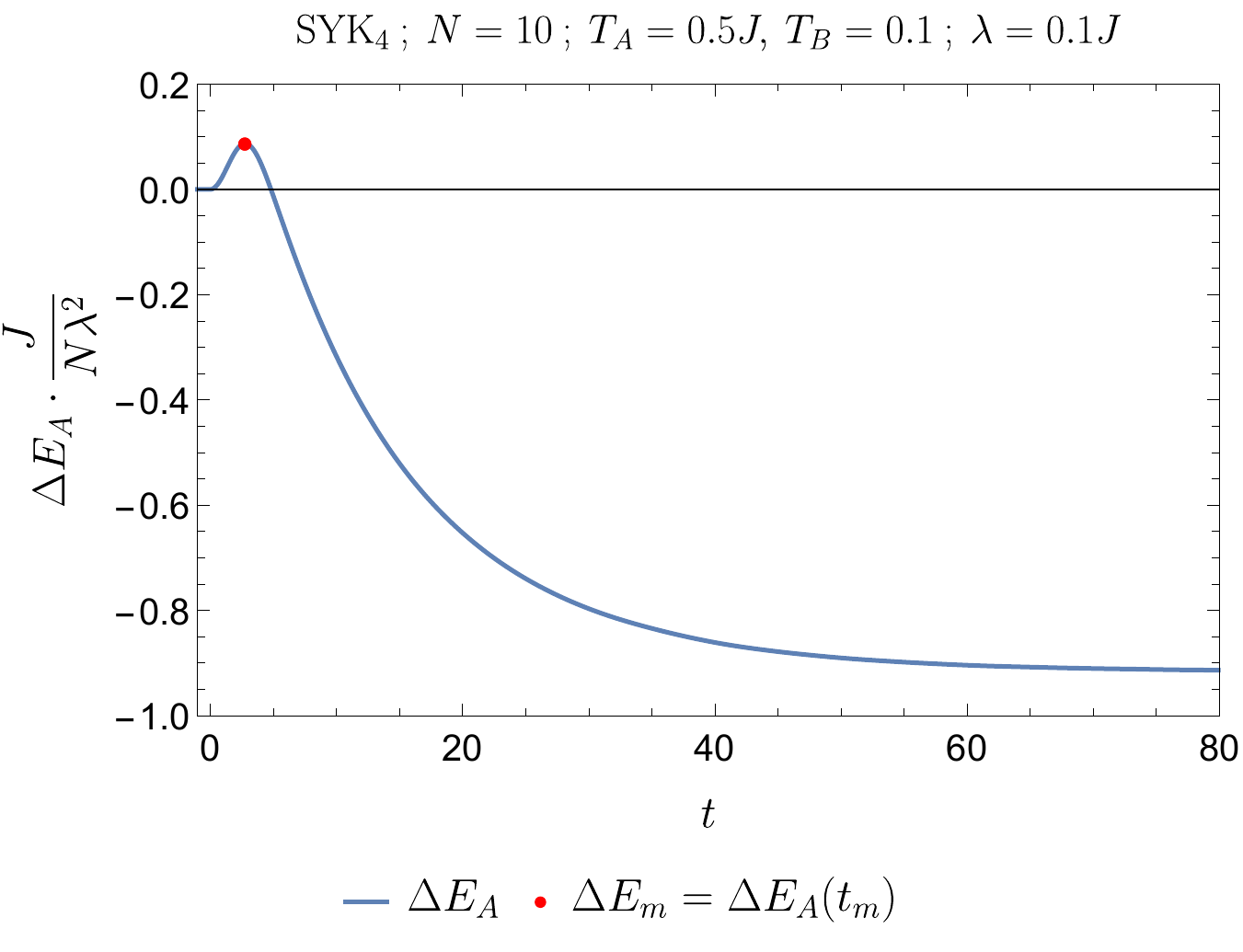}
\caption{Normalized change of the energy 
of the hotter system $A$ $(\Delta E_A= \Tr ((\rho_A(t)-\rho_{T_A}) H_A)$ as a function of time. At short times it increases counter to intuition.  Majorana SYK${}_4$ in exact diagonalization averaged over $R=100$ realizations with parameters of both systems on top of the plot. Red dot marks the bump that is reached at time $t_m$ and has a height $E_m$ relative to the initial energy.
}
\label{fig:E1_SYK_twoRegimes_bump}
\end{figure}

Fig.~\ref{fig:E1_SYK_twoRegimes_bump} shows the classically unexpected rise in energy in system A directly following the cooling quench with $T_A> T_B$ found in \cite{Zhang2019Evaporation,Almheiri2019Universal}.
We shall now show that even though $E_A$ initially increases, there is no energy flux from the cold reservoir to the hot system. The energy increase instead follows  from the energy contribution of the interaction Hamiltonian solely but it is nevertheless a real modification of energy, as a subsequent decoupling of A and B shows. At the moment of decoupling work must be performed on the combined system-reservoir as we shall show.

The above conclusions follows from the following observations in 
SYK systems:

\begin{enumerate}
    \item Directly following the quench, the system-energy $E_A(t)$ and the reservoir-energy $E_B(t)$ {\em both} grow 
    (Fig.\ref{fig:EAll_SYK_twoRegimes_bump}).
    The fact that there is no net energy flow from cold to hot means the energy must come from somewhere else.
    
\begin{figure}[!h]
\includegraphics[width=0.5\textwidth]{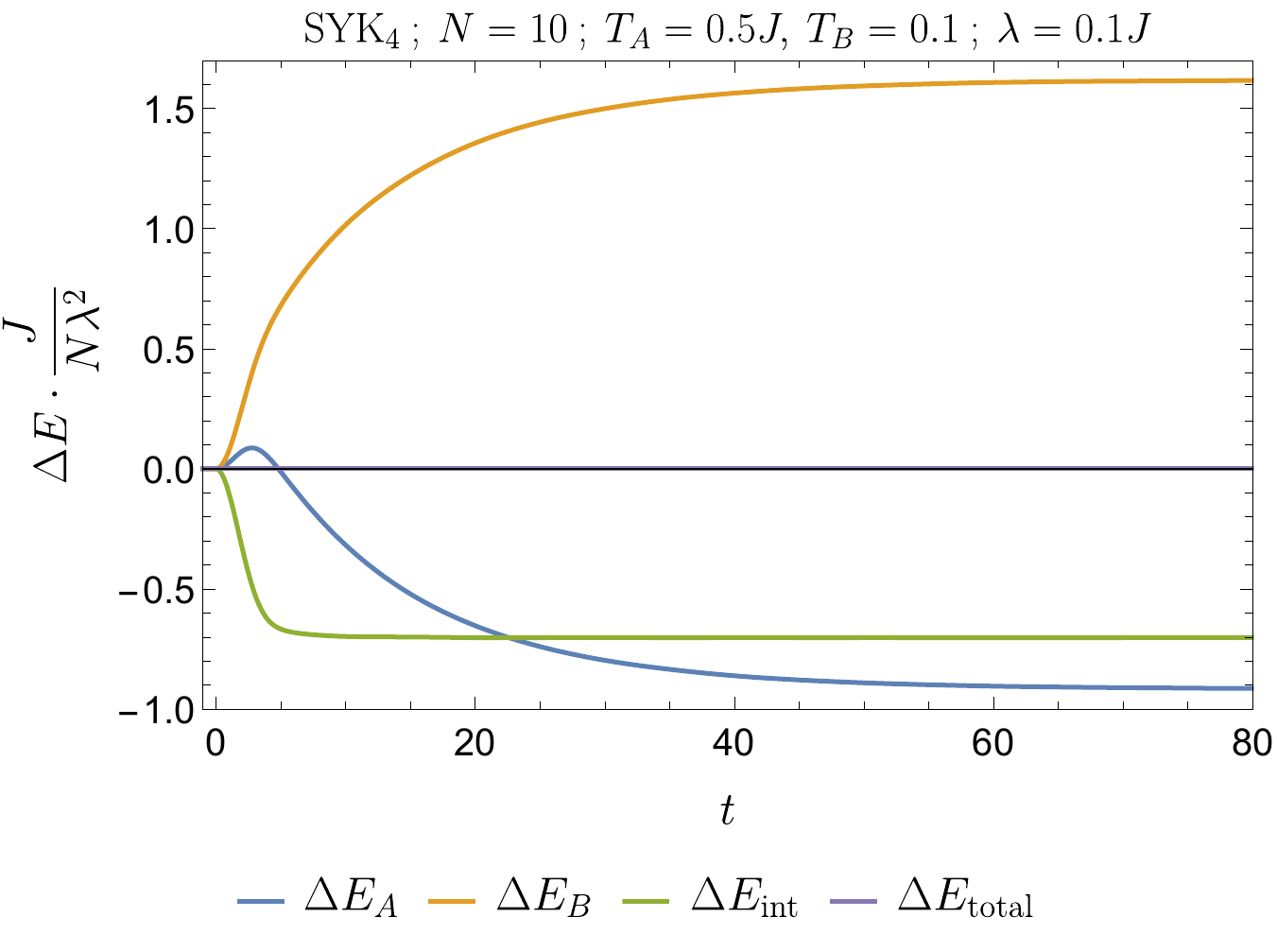}
\caption{Normalized change of energy of the two subsystem $A$, $B$, the interaction energy $\Delta E_{\text{int}}= \Tr ((\rho(t)-\rho_0) H_{\text{int}} \}$ and the total energy $E_{\text{total}}$ as a function of time. Directly following the quench both the system-energy $E_A(t)$ and the reservoir-energy $E_B(t)$ grow, whereas the interaction energy $E_{\text{int}}(t)$ decreases. The sum vanishes as must be as no energy is put into the combined system/reservoir. {Majorana SYK in exact diagonalization} averaged over $R=100$  realizations with parameters of both systems on top of the plot.}
\label{fig:EAll_SYK_twoRegimes_bump}
\end{figure}

    \item The total Hamiltonian $H_{\text{total}}=H_A+H_B+\theta(t)H_{\text{int}}$ contains a third contribution $H_{\text{int}}$. Its contribution to the energy is negative (Fig.\ref{fig:EAll_SYK_twoRegimes_bump}).

    \item The change in the expectation value in the total Hamiltonian is nevertheless readily computed to vanish. 
    \begin{align}\label{eq:dHfdt}
        \frac{d}{dt}\langle H_{\text{total}} \rangle = i\langle [H_{\text{total}},H_{\text{total}}]\rangle + \delta(t)\langle H_{\text{int}}\rangle
    \end{align} 
    The first term vanishes trivially. 
When
    $\langle H_{int}\rangle (0)=0$ as well, as is the case in all the systems we study, then $\langle{H_{total}}\rangle$ is constant in time. The ``binding''-energy from $E_{\text{bind}}=-E_{\text{int}}(t)=\Tr(H_{\text{int}}\rho(t))$ thus 
    completely accounts for the rise in both $E_A(t)$ and $E_B(t)$.
    
    \item More precisely, for $E_A(t)$ to correspond to a {\em measurable}  
    energy change  (in the sense of commuting with the Hamiltonian) 
    one should decouple the system from the reservoir with a second quench at a finite time $t_f$ later, as in the standard two-point measurement protocol in quantum thermodynamics \cite{Goold_2016,deffner2019quantum,Popovic2021Thermodynamics}. 
    Then $H_A$ commutes again with the full Hamiltonian for $t>t_f$.\footnote{Formally, if one does not decouple, the eigenstates of $H_{\text{tot}}$ are no longer localized within $A$ or $B$, and one cannot really say that the expectation value of $H_A$ is the energy of the sub-system $A$. 
    The expectation value of $H_A$ nevertheless comes the closest and is therefore what is conveniently called the energy of this subsystem. 
    } 
    In other words, as in our previous article \cite{gnezdilov2021information}, one considers the two-quench protocol $H_{\text{total}}=H_A+H_B+(\theta(t)-\theta(t_f))H_{\text{int}}$.
    Computing the change in total energy, one clearly sees that the energy that must now be supplied equals the binding-energy $E_{\text{bind}}=-E_{\text{int}}(t_f)$.
    \begin{align}
        \frac{d}{dt}\langle H_{\text{total}}\rangle = -\delta(t_f)\langle H_{\text{int}}\rangle~.
    \end{align}
    Choosing $t_f$ during the initial time period where both $E_A$ and $E_B$ increase, one concludes that for a two-point measurement protocol of such short duration the total energy in the system has increased.
    In particular there are initial configurations of $T_A,T_B$ 
    where the final equilibrium temperature after such a short-time two measurement protocol is larger than both $T_A$ and $T_B$; see Fig.\ref{fig:E_turning_off}.
    The decoupling 
    quench must therefore perform work on the system.
\begin{figure}[!t]
\centering
\includegraphics[width=0.5\textwidth]{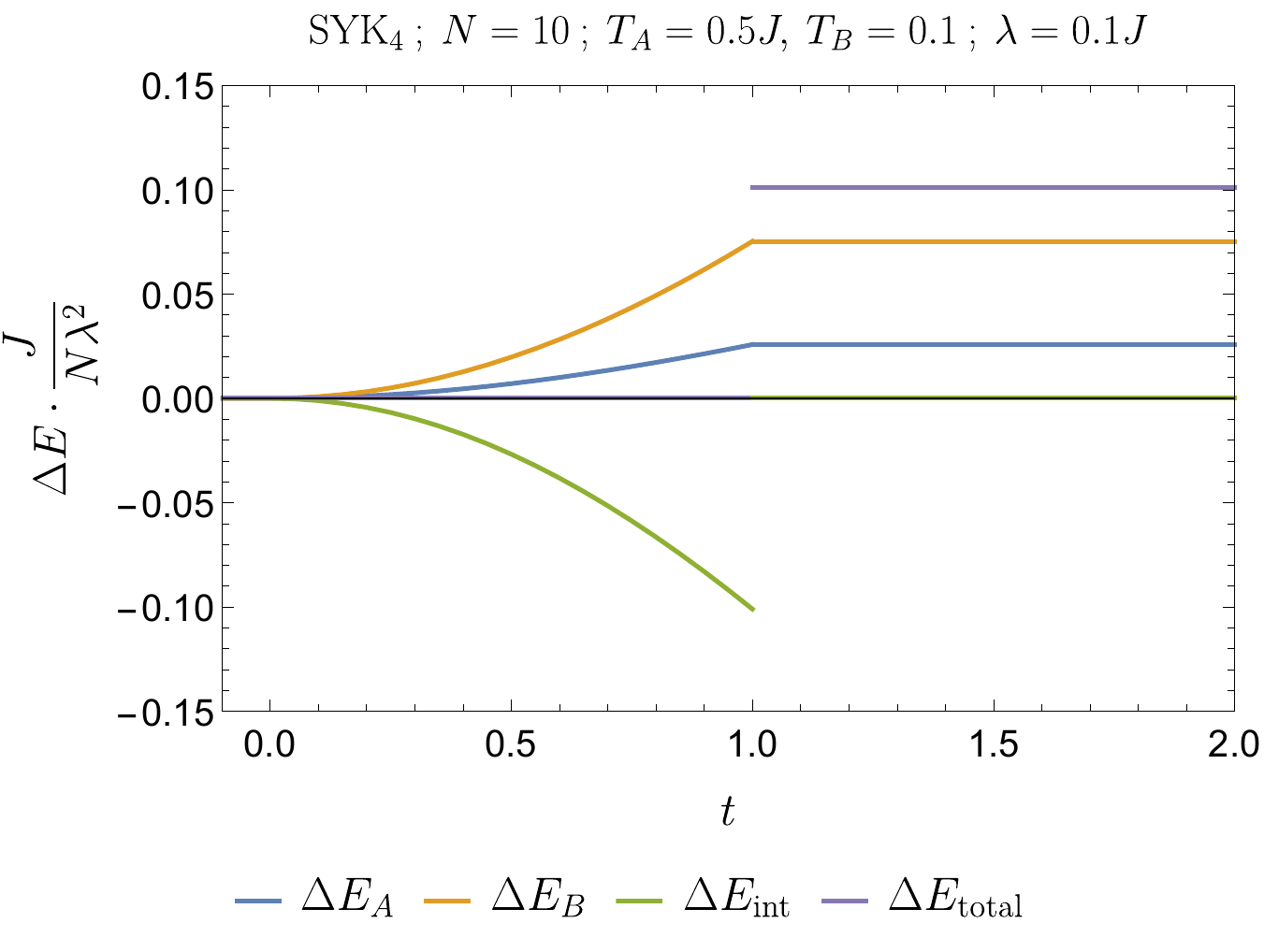}
\caption{Normalized change of energy of the two subsystem $A$, $B$, the interaction energy $\Delta E_{\text{int}}= \Tr ((\rho(t)-\rho_0) H_{\text{int}} \}$ and the total energy $E_{\text{total}}$ as a function of time in a two-quench protocal with the interaction turned of at $t_f=1$. At $t_f$ the change in total energy shows the energy supplied to the system which exactly equals $E_{\text{int}}$. {Majorana SYK in exact diagonalization} averaged over $R=100$ realizations with parameters of both systems are on top of the plot.}
\label{fig:E_turning_off}
\end{figure}    
    
    \item
    {In general, since the whole system $AB$ is closed, the total change in the energy of each subsystem, $A$ or $B$, can be due to two components, the contribution from/debit to the ``binding''-energy and the  thermal exchange between $A$ and $B$:
    \begin{subequations}
    \begin{gather}
        \Delta E_A = \Delta E_{A,bind} + \Delta E_{B~to~A} \label{eq:EAeqEAbind_EBtoA}
        \\
        \Delta E_B = \Delta E_{B,bind} - \Delta E_{B~to~A}. \label{eq:EBeqEBbind_EBtoA}
    \end{gather}
    \end{subequations}
    We can estimate the binding energy for each subsystem $A, B$ with respective initial temperatures $T_A\neq T_B$ separately from the interaction energy of a second quench experiment with an equal temperature setup $E_{\alpha,bind}\approx -\frac{1}{2}E_{\text{int}}(T_A=T_B=T_\alpha)$, i.e. we determine $E_{A,{bind}}$ from a quench set-up where both system and reservoir have initial temperature $T_A$, and $E_{B,\text{bind}}$ from a quench set-up where both system and reservoir have initial temperature $T_B$. 
    {Using this estimate in the quenched cooling set-up with different temperatures that are not too different} we can numerically compute the thermal flux from $B$ into $A$ as
    \begin{gather}
      \Phi_A=\frac{d}{dt}{E}_{B~to~A}=\frac{1}{2}\left(\frac{d}{dt}{E}_A-\frac{d}{dt}{E}_B \right) -\frac{1}{2}\left(\frac{d}{dt}{E}_{A,bind}-\frac{d}{dt}{E}_{B,bind} \right) .  
      \label{eq:fluxBtoA}
    \end{gather}
    The flux $\Phi_A$ is always negative and at early times it is subdominant to the binding energy Fig. \ref{fig:Energy_Flux_AtoB}. This proves that even when $E_A$ increases initially, the  energy flux/heat transport is nevertheless always from the hot system A to the cold reservoir B and the supplied energy for the increase comes solely from the binding-energy or the outside when decoupling $A$ and $B$.
    }

\begin{figure}[t] 
\centering
\includegraphics[width=0.5\textwidth]{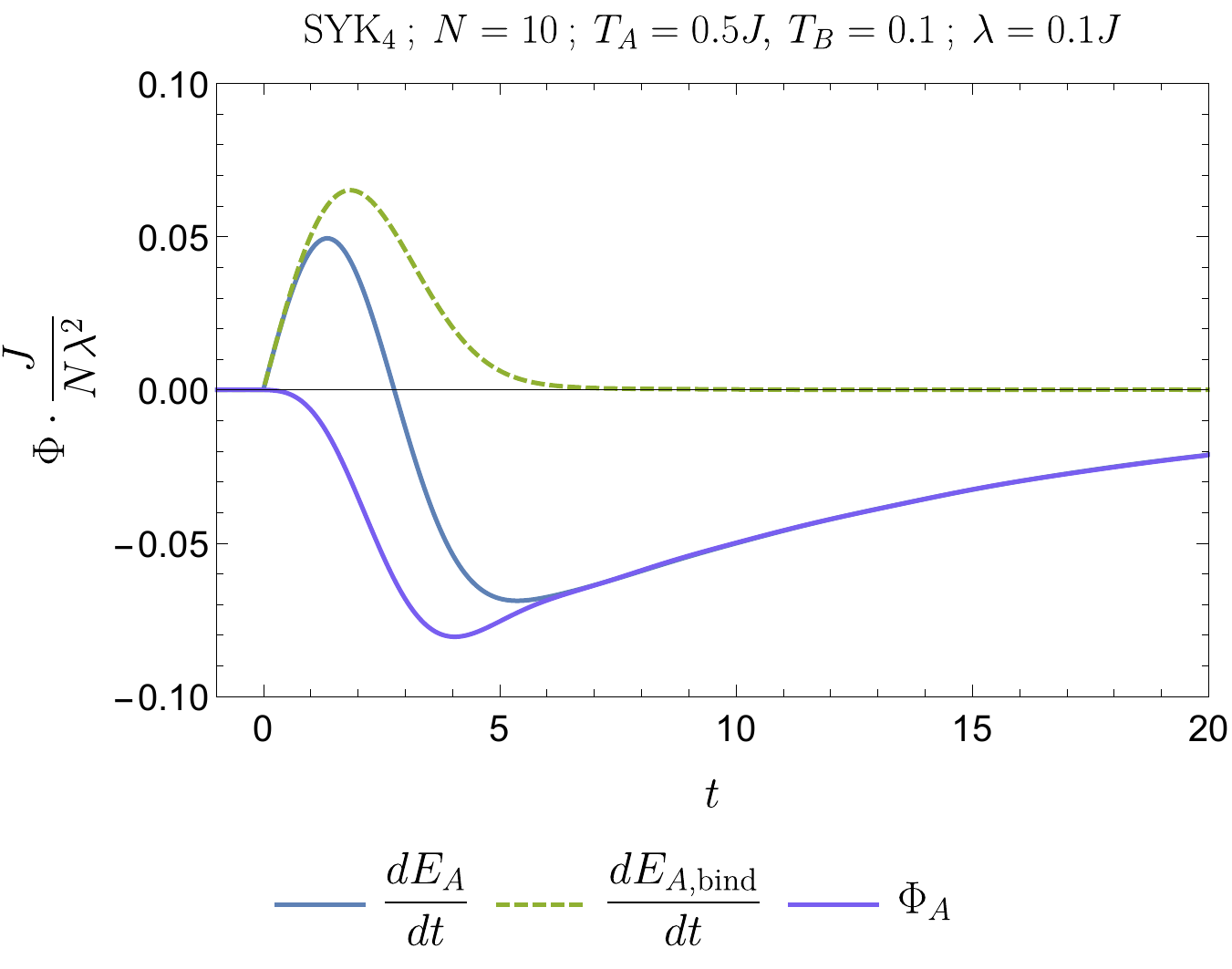}
\caption{Time derivatives of the energy $E_A$ of subsystem $A$; time derivative of an estimate of binding energy contribution $E_{A,\text{bind}}$ from considering an equal temperature quench $(T_A=T_B=0.5J)$, and the resultant thermal flux {from cold reservoir $B$ to hot system $A$}. The flux is always negative and always flows from hot to cold. 
Majorana SYK in exact diagonalization averaged over $R=100$ realizations with parameters of both systems  on top of the plot.
}
\label{fig:Energy_Flux_AtoB}
\end{figure}   
\end{enumerate}

\subsection{Energy rise driven by {quantum correlations}}
As previewed in the introduction the quantity that controls this rise in energy $E_A$ from the contribution of the ``binding''-energy to the combined system-reservoir is the von Neumann {entropy of the reduced density matrix of} system $A$: {$\rho_A(t)=\Tr_B\rho(t)$ }. To see this, consider the relative entropy between $\rho_A(t)$ and the initial thermal density matrix
\begin{align}
    D(\rho_A(t)||\rho_{T_A}) = \Tr(\rho_A(t)\ln\rho_A(t))-\Tr(\rho_A(t)\ln\rho_{T_A})~.
\end{align}

Substituting that $\rho_{T_A}=\frac{1}{Z_A}e^{-\hat{H}_{A}/T_A}$ one immediately has
\begin{align}
\label{eq:RelEntWithThermal}
    T_A D(\rho_A(t)||\rho_{T_A})+T_A S_{\vN,A}(t) = E_A(t) -F_A~.
\end{align}
where $F_A=-\ln Z_A = E_{A}(0)-T_AS_A(0)$ is the free energy of the initial thermal state.
The time-dependent terms form the definition of the 
{\em information free energy} 
\begin{align}
    {\cal F}(t:{T_A}) = E_A(t)-T_A  S_{\vN,A}(t) = F_A + T_A D(\rho_A(t)||\rho_{T_A}).
\end{align}
It encodes the energy-available-for-work and its full counting statistics in open quantum systems that decohere due to their interaction with the environment. The loss of information due to decoherence and decorrelation costs work according the Landauer's principle and the information free energy accounts for that \cite{Goold_2016,deffner2019quantum}.

\begin{figure}[t!]
  \includegraphics[width=0.7\textwidth]{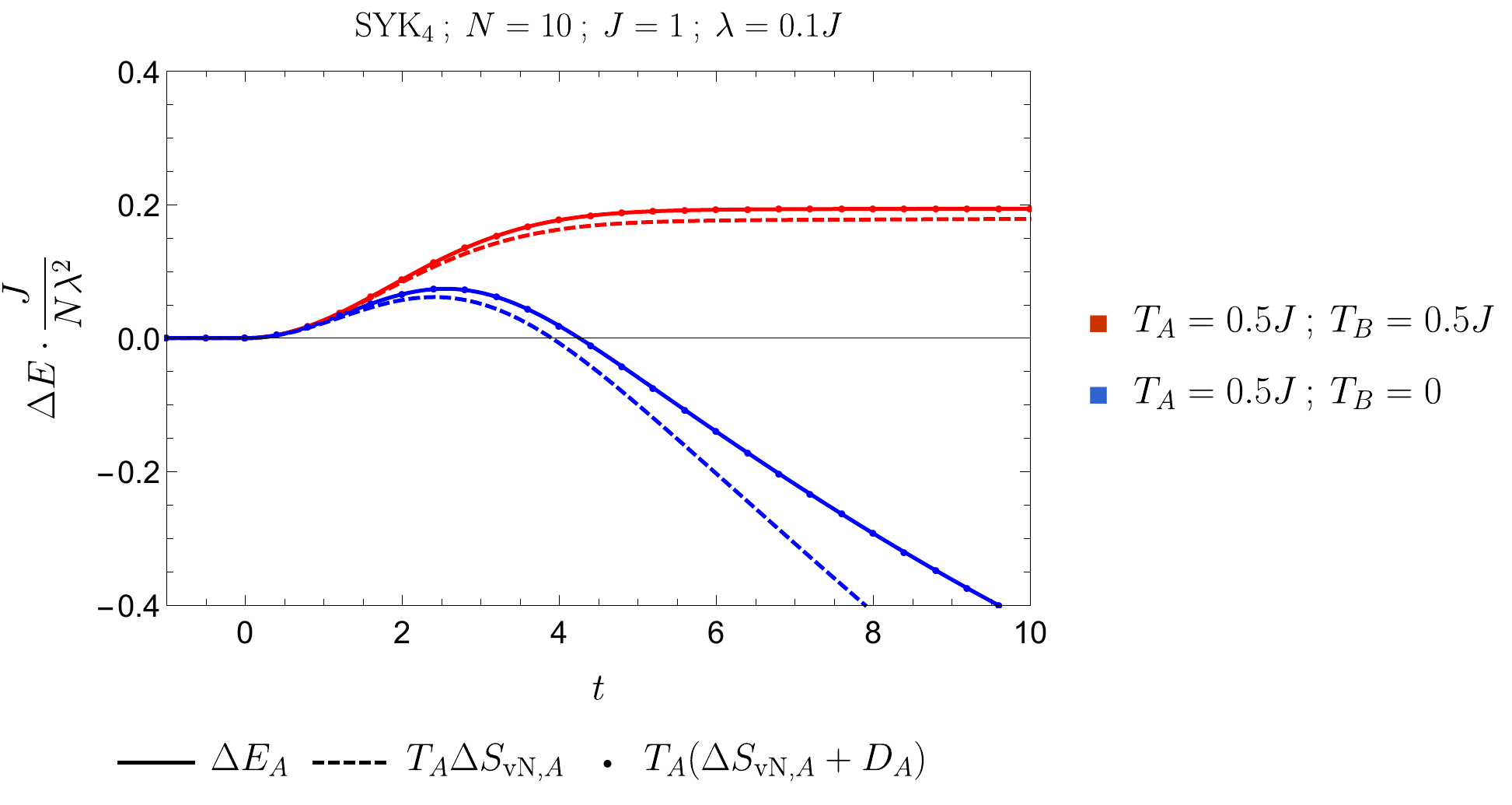}
  \caption{{The 
  energy $E_A$ is verified to equal 
  the sum
  of the von Neumann entropy $\Delta S_{\text{vN},A}$ times the initial temperature $T_A$ and the relative entropy $D_A\equiv D(\rho_A(t)||\rho_{T_A})$. The initial rise in the energy in particular is controlled by the initial rise in the von Neumann entropy.}
  This persists when the reservoir is in the groundstate $T=0$ and at equal system-and-reservoir temperature $T_A=T_B$ pointing to entanglement as cause of the rise in von-Neumann entropy. Data from  Majorana SYK in exact diagonalization averaged over $R=100$  realizations with parameters on top of the plot. 
  }
  \label{fig:OhanesjansLaw}
\end{figure}

The change in energy of system $A$ after the quench directly follows from Eq.\eqref{eq:RelEntWithThermal} 
and immediately brings us to Eq.(\ref{eq:1a}). 
\begin{align}
  \Delta E_A(t)=E_A(t)-E_A(0)= T_A\Delta S_{\vN,A}(t)+T_A D(\rho_A(t)||\rho_{T_A}),
  \nonumber
\end{align}
{and using the semi-positive definiteness of the relative entropy} Eq.\eqref{eq:1}
\begin{align}
    \Delta E_A(t) \geq T_A\Delta S_{\vN,A}.
    \nonumber
\end{align}
Both the equality and the inequality are readily observed in exact diagonalization of Majorana SYK models, see Fig.\ref{fig:OhanesjansLaw}.

\bigskip
\bigskip

Two important remarks can be made:
\begin{enumerate}
\item As the relative entropy is very small at early times the initial rise in energy is completely determined by the rise in the von-Neumann entropy.\footnote{Strictly speaking fine tuned initial conditions 
can exist where the von-Neumann entropy decreases, but decreases so little that the small rise in relative entropy nevertheless results in an energy increase in the hotter system. 
}

\item This 
rise is even present when the reservoir $B$ is at $T_B=0$, as well as when the system and reservoir are at equal $T$ (Fig.\ref{fig:OhanesjansLaw}).  This unambiguously points to the growth of quantum entanglement as the contributing factor to the rise in the von-Neumann entropy; (see also \cite{gnezdilov2021information}).
\end{enumerate}

Given that it is the von Neumann entropy growth that controls the early time dynamics between the two subsystem, it is natural to also consider the evolution of mutual information between the two:\footnote{When the system and the reservoir have equal $T$, then
\begin{align}
    \Delta E_A(t)+\Delta E_B(t) 
    &=T\Delta I(A:B) +D(\rho_A(t)||\rho_T)+D(\rho_B(t)||\rho_T)) ~.
\end{align}
since $\Delta S_{A\union B}(t)=\Delta S_{\text{total}}(t)=0$ due to unitary evolution of the combined system-reservoir combination as a whole. In the early time regime where the relative entropies are very small, the combined energy change in $A$ and $B$, equal to work needed at the moment of a decoupling quench, is then equal to the mutual information. This was first pointed out in \cite{Groisman_2005} where it was shown that the minimum amount of noise to decorrelate two systems equals the mutual information. By Landauer's principle this is then also the minimal amount of work. Note, however, that the energy increase here is not directly related to decorrelation between $A$ and $B$.}
\begin{align}
    I(A:B,t)=S_{\text{vN},A}(t)+S_{\text{vN},B}(t)-S_{\text{vN},A\union B}(t)~,
\end{align}
where  $S_{\text{vN},A\union B} = - \mathrm{Tr}_{A,B} \, \rho_{A\union B} \ln \rho_{A\union B}$ with $\rho_{A\union B}$ being the density matrix of the full system.
It displays two qualitatively distinct regimes: an
initial polynomial increase followed by an exponentially decaying approach to equilibrium.
Qualitatively, the early time $(t<t_{m})$ behaviour of the mutual information resembles the results reported in \cite{Deffner_2020,Deffner_2021} where mutual information was used as a better measure of quantum scrambling, compared to the OTOC. In particular these articles prove that $I(A:B)$ bounds the OTOC from above. This supports our deduction above that the initial energy increase is caused by quantum {correlation- and/or} entanglement-growth and scrambling. 
Note that the OTOC of operators between two quenched quantum dots depends on the initial state and interaction between the two dots, hence the early time polynomial increase in our setup. This should not be confused with the exponential growth of OTOC within a single SYK dot, which is driven by strong entanglement. 
The articles \cite{Deffner_2020,Deffner_2021} also emphasize the role of decoherence in addition to scrambling. It would be interesting to dissect and analyze their interplay further
 but we leave this for the future.

\begin{figure}[h]
  \includegraphics[width=0.5\textwidth]{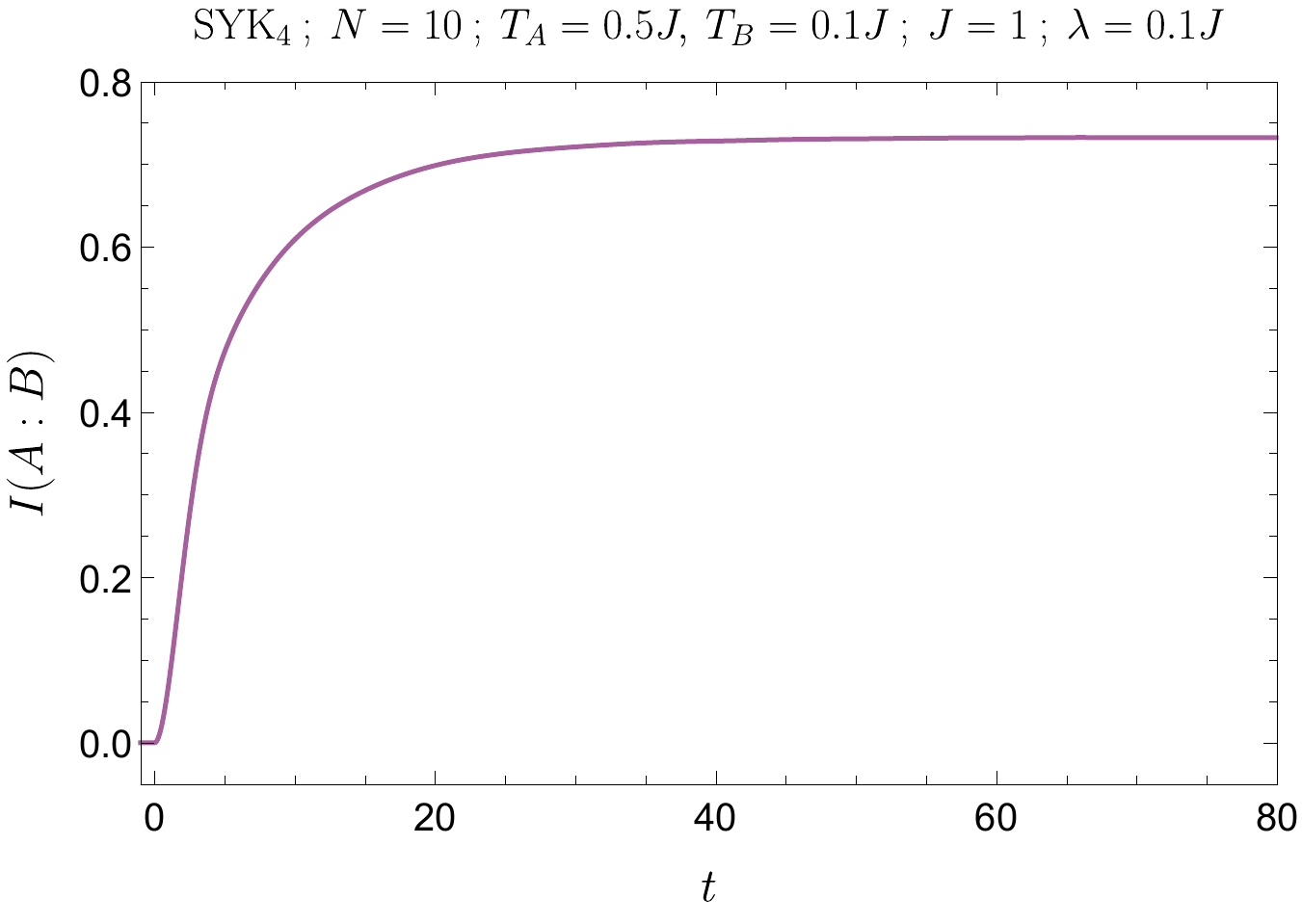}
\caption{Growth of the mutual information between subsystems A and B.}
  \label{fig:mutual_information}
\end{figure}

\section{The transition from quantum to classical cooling}

\begin{figure}[t!]
    \includegraphics[width=0.6\textwidth]{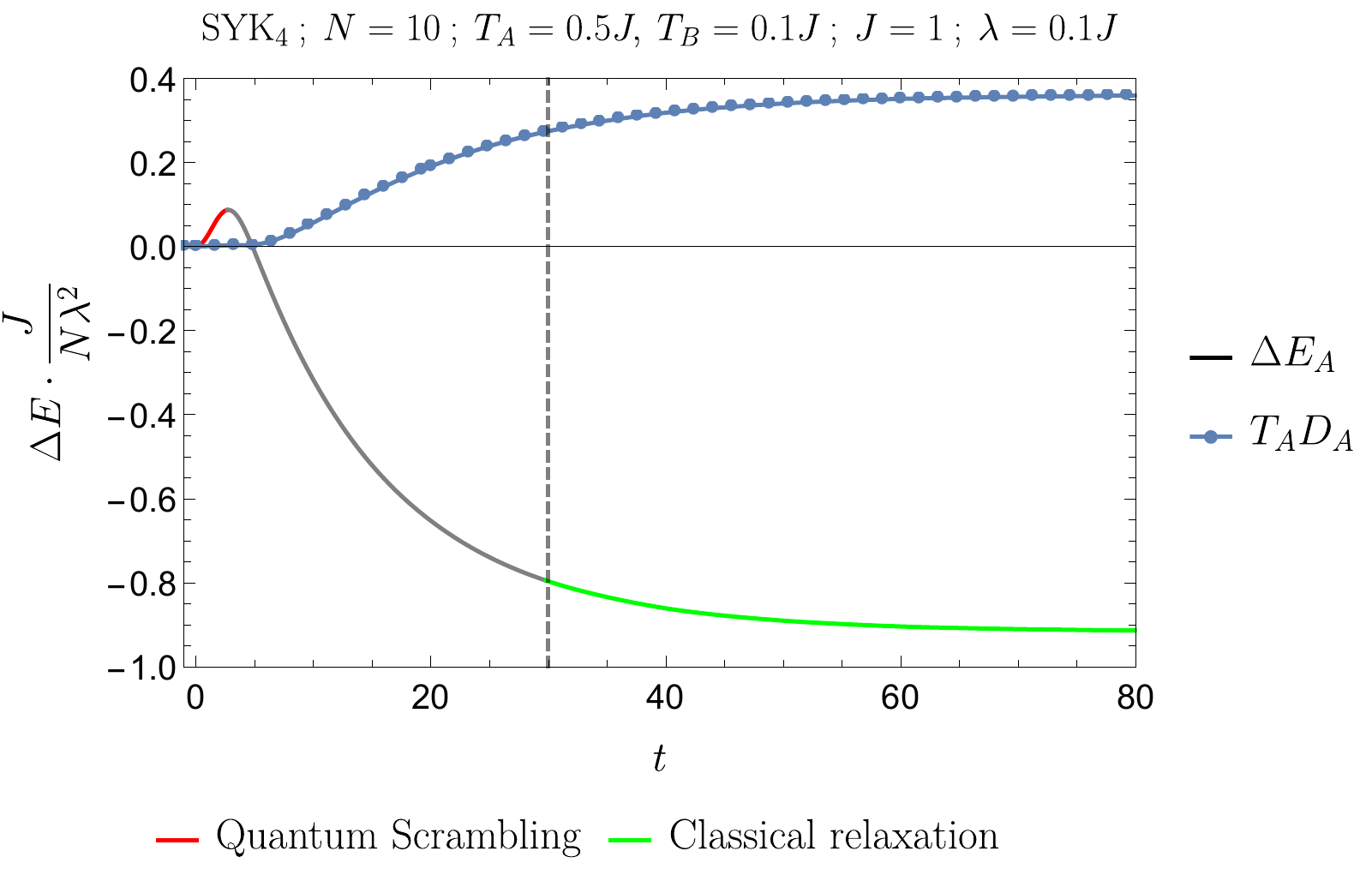}
    \caption{The generic contact quench is characterized by an early time quantum scrambling dominated regime (red) that transitions to a regime exhibiting conventional classical relaxation (green). The transitions between these regimes are not sharp, but roughly indicated by the top of the initial energy bump and the saturation of the relative entropy, where the final density matrix has become approximately thermal. 
    }
    \label{fig:qu-vs-class} 
\end{figure}

At late times after the quench the system behaves fully as expected in that the energy of the hotter system exponentially decreases until it equilibrates.
Given that the initial rise of energy is controlled by the rise in entanglement driven von-Neumann entropy, there are two clear  regimes: this initial rise and the late time relaxation (Fig.\ref{fig:qu-vs-class}). For the specific case of the quenched cooling two SYK dots, one can use the fact that large $N$ SYK is exactly solvable to make analytic estimates for both these regimes as well as the intermediate regime and the long time hydrodynamic tails which eventually change the relaxation to equilibrium from exponential to power law \cite{Almheiri2019Universal}. 

Here we ask a different question. Having argued that the
initial rise is generically universally controlled by the rising 
quantum correlation
contribution to the von-Neumann entropy, under what circumstances does the expected classical physics emerge, where heat immediately flows from hot to cold? The quantum 
correlation- and/or
entanglement-growth is always present (except if the  full system is purely classical where  all the terms in the full Hamiltonian, including the coupling term, commute with each other).  This can therefore only happen in circumstances where the ``classical''
relaxation overwhelms the quantum growth. Or more precisely, knowing  that 
\begin{align}
    \Delta E_A(t) \geq T_A\Delta S_{\vN,A}(t), \nonumber
\end{align}
this transition can only happen if the ``classical'' thermal contribution to the von-Neumann entropy dominates over the entanglement contribution to the von-Neumann entropy already at the earliest possible time.
From the atomic statistical mechanics underpinning of classical thermodynamics we know that this must happen when we have a theory with well defined particles with suppressed quantum correlations. This should be the case at high temperatures (weak coupling) and low densities. 

\begin{figure}[t!]
    \includegraphics[width=0.49\textwidth]{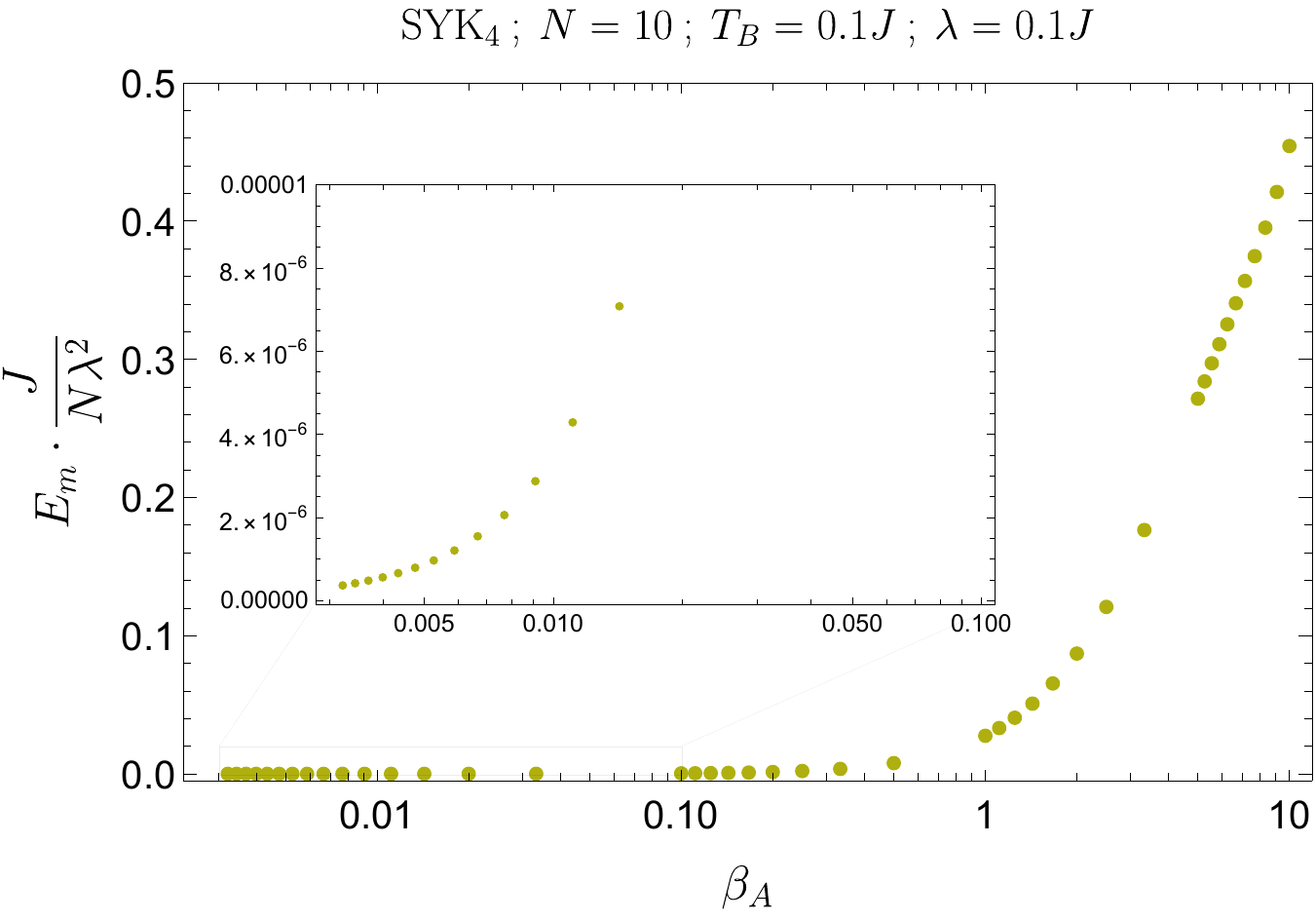} 
    \includegraphics[width=0.49\textwidth]{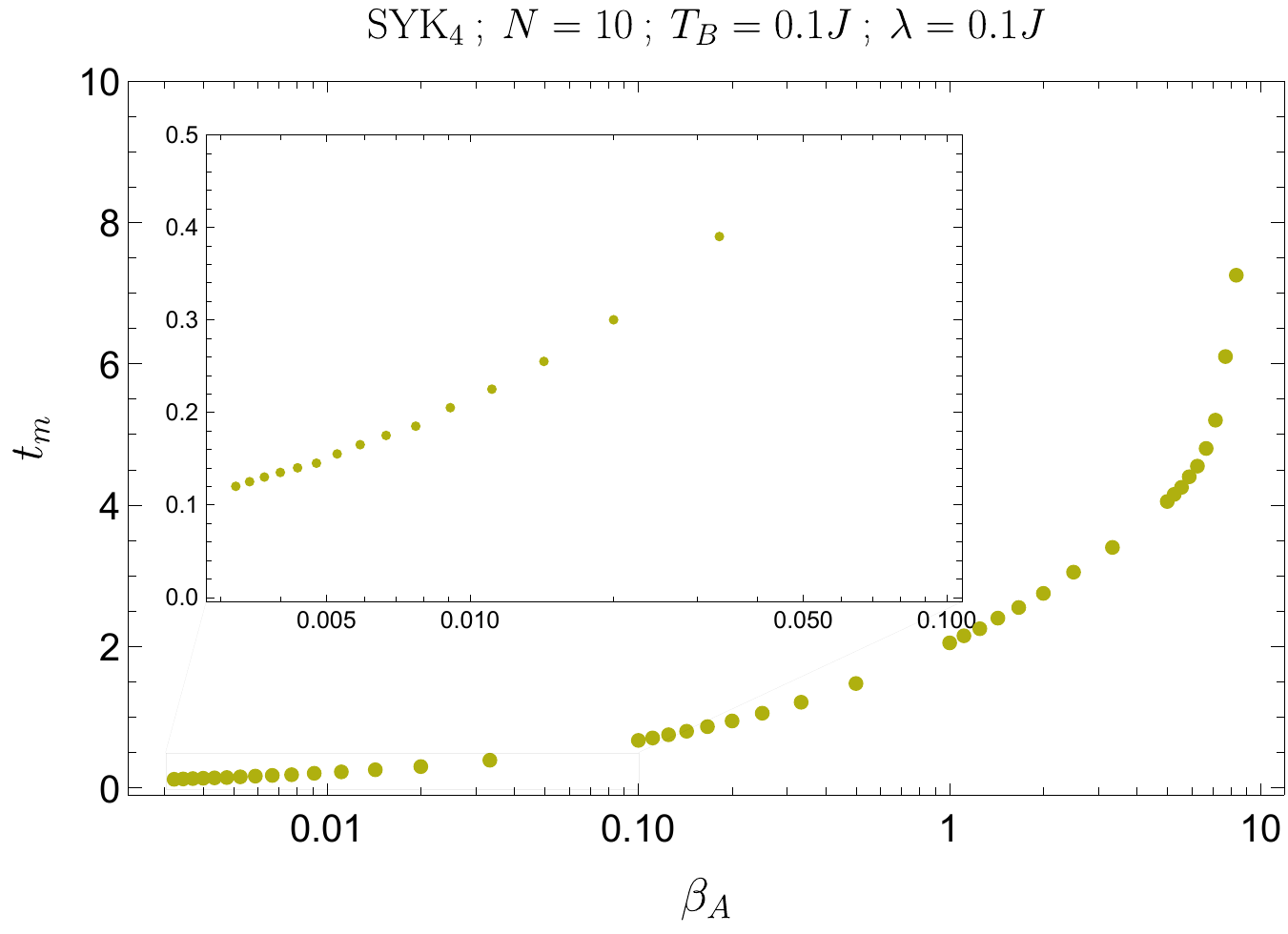}
    \\[.1in]
    \includegraphics[width=1\textwidth]{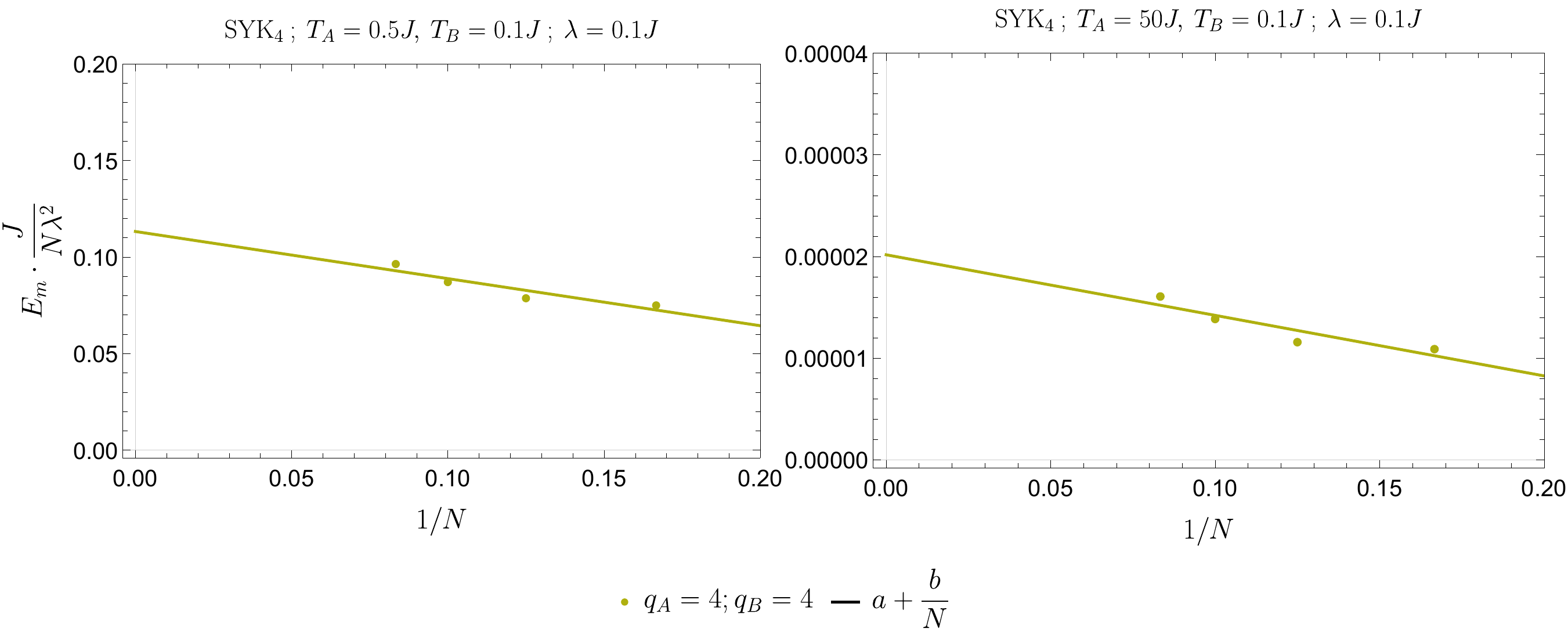}
    \caption{Quenched cooling of two SYK$_4$ dots. {\bf Top:} Height $E_m$ of the energy bump (left) and time $t_m$ of the bump (right) for various initial temperatures $T_A=1/\beta_A$.
    {\bf Bottom:} Height $E_m$ of the energy bump roughly extrapolated to larger $N$ for two different initial temperatures $\beta_A$. The height stays finite in this thermodynamic limit, indicated by $a>0$. 
    Combining the top and the bottom, the 
    initial rise in the hotter system energy $E_A$ seems to persist for any finite $T_A$ and infinite $N$.
    }
    \label{fig:SYKbumpheight_log}
\end{figure}

However, when we study the high $T$ ($T_A, T_B \gg J^2$ and $T_A \gg T_B$) regime in quenched cooling two SYK$_4$-dots, this disappearance of the initial rise and a transition to immediate classical energy flow from hot to cold is not seen to
emerge. 
This is even so when we extrapolate our finite size exact diagonalization result to the thermodynamic limit $(N \rightarrow \infty)$ {(with the assumption that the finite $N$ studies do capture the appropriate large $N$ behavior)}. Fig.~\ref{fig:SYKbumpheight_log} shows the height of the energy bump $E_{m}=E_{max}-E(t=0)$ per particle $(E_{m}/N)$ in the Majorana SYK$_4$ model directly before it starts to decrease as a function of the temperature $T_A$. Any finite $N$ system will always contain quantum signatures and the classical behavior need only emerge in a thermodynamic limit. Numerics directly gives away that $E_m$ has a leading scaling with $N$.
Dividing this overall scaling out, a rough extrapolation to  $N=\infty$ nevertheless shows that a positive energy bump remains.\footnote{This turns out to also be true for SYK$_2$ models. Though within the random ensemble of SYK$_2$  couplings, there are empirically always realizations for which the energy $E_A$ does decrease instantaneously. 
}

\begin{figure}
    \includegraphics[width=0.49\textwidth]{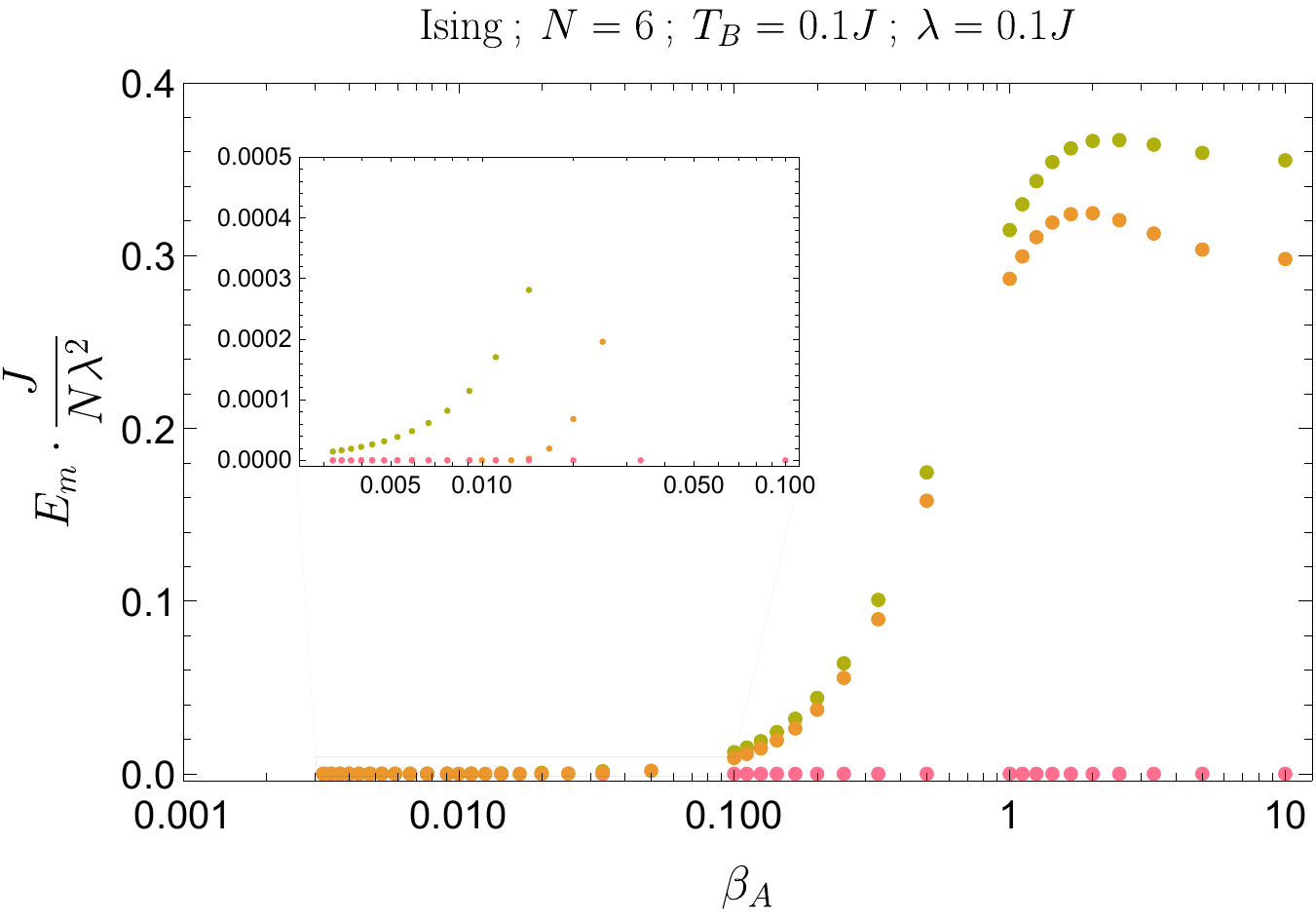}
    \includegraphics[width=0.49\textwidth]{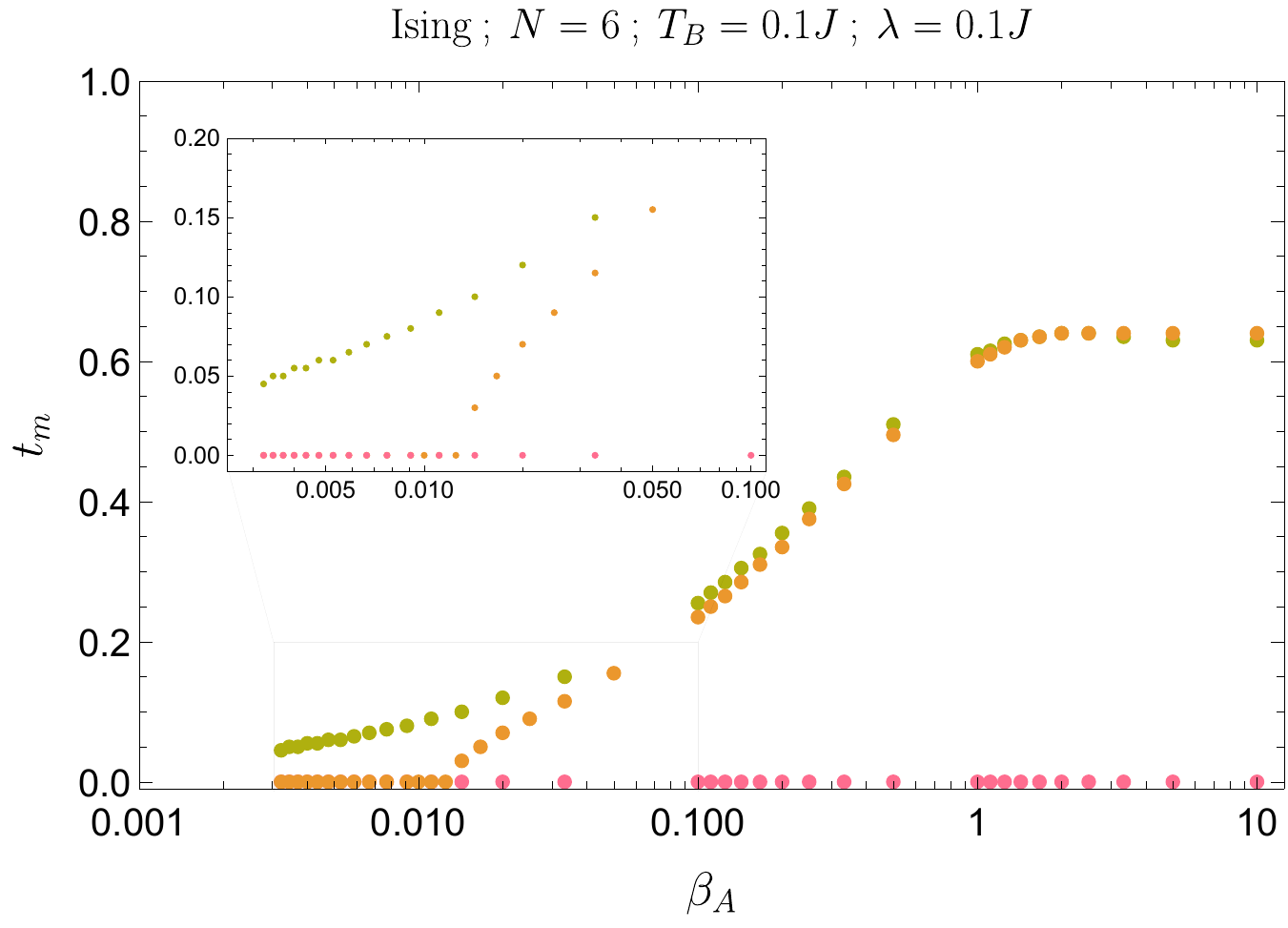}
    \\[.1in]
    \includegraphics[width=1\textwidth]{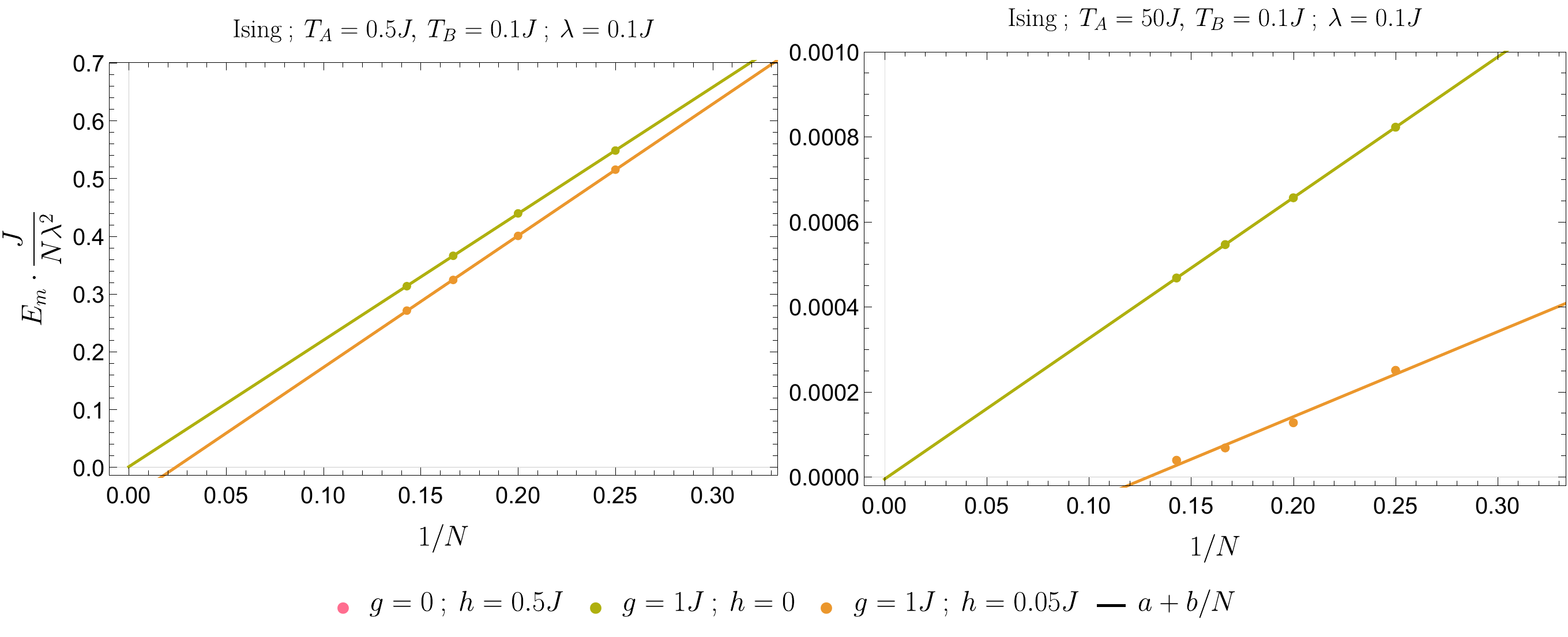}
    \caption{Quenched cooling in two Ising half lines. 
     {\bf Top:} Height $E_m$ of the energy bump (left) and time $t_m$ of the bump (right) in  for various parameter choices. {\bf Bottom:} Height $E_m$ of the energy bump extrapolated to larger $N$ for various initial temperatures $T_A=1/\beta_A$. For each initial temperature there is a finite extrapolated value of $N$ for which the bump disappears $(a\leq 0)$ and the system will cool instantaneously upon contact. The higher the initial temperature, the lower is this value of $N$.}
    \label{fig:IsingBumpheight}
\end{figure}

To try to find the crossover to expected classical behavior where the energy rise in the hot system is absent, we change
the quenched cooling set-up from two SYK quantum dots to two mixed field Ising half-lines Eq. \eqref{eq:H_Ising} with a tunneling interaction at the end point of each line Eq.\eqref{eq:H_Ising_Interaction}. 
Both at the free $g=0,h=0$ \cite{Doyon:2014qsa} and at the conformal fixed point $g=1, h=0$ in the continuum (thermodynamic) limit one can use conformal field theory techniques to study this type of quenched cooling \cite{Bernard:2012je,Bernard:2013aru,Bhaseen:2013ypa}. Then one indeed finds that there is no initial energy rise, but the energy starts to flow instantaneously from hot to cold.
As is well known by now, in the regime $h=0$ 
the late time behavior of the two subsystems, if isolated, is controlled by the large number of conserved charges and an associated generalized hydrodynamical relaxation towards a generalized Gibbs ensemble \cite{DeLuca:2013laa,2020GHDLectures}. The presence of the coupling term $\lambda$ makes the full system not integrable.

\begin{figure}[!t]
\includegraphics[width=0.5\textwidth]{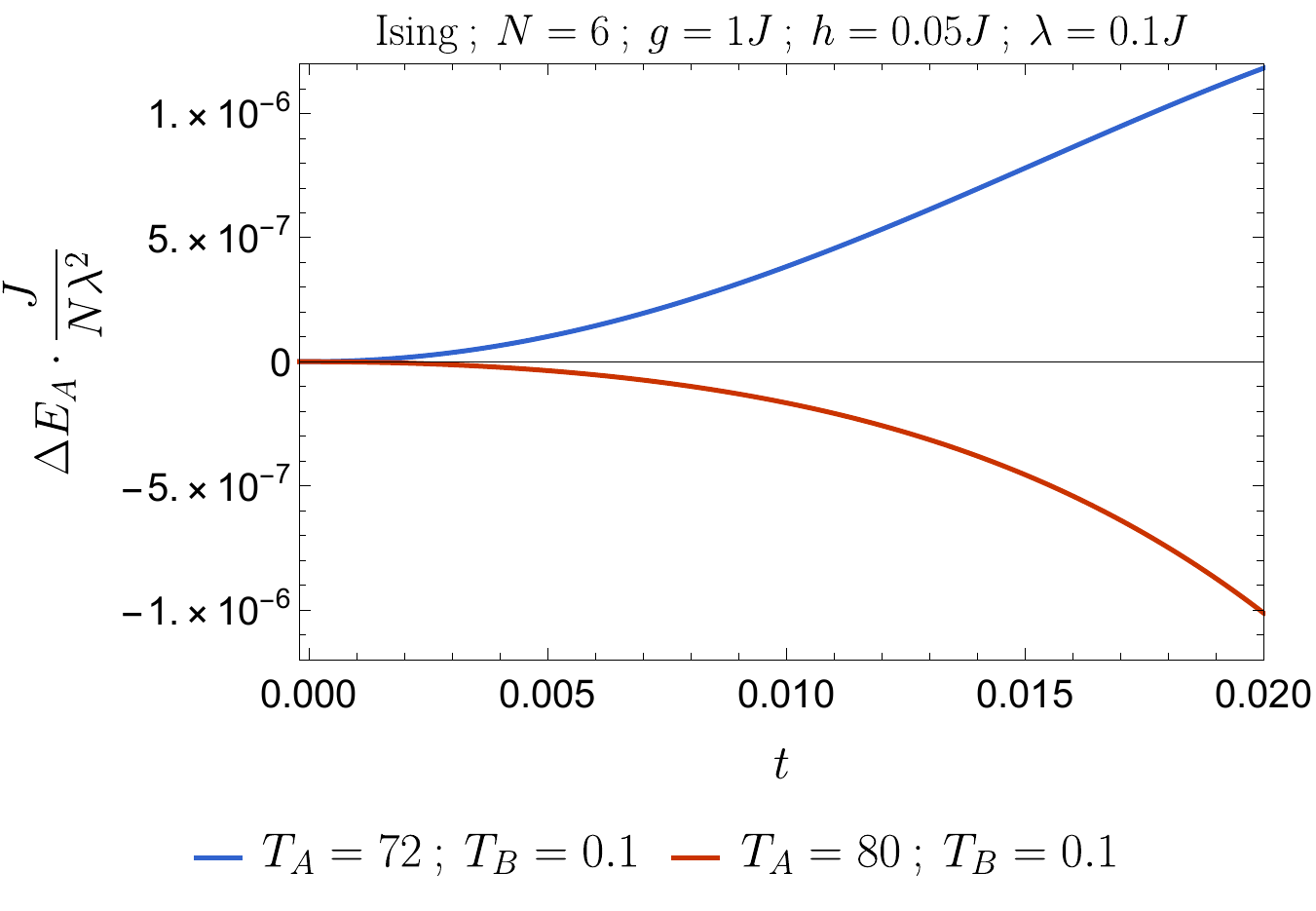} 
\caption{Quenched cooling in two Ising half lines. For $T<T_c\simeq 77.845 J$ 
one still observes the 
initial rise in the hotter system $A$, but for $T>T_c$ one transitions to a regime where classical intuition is restored and the system cools 
instantaneously upon contact.}
\label{fig:fixedptIsing}
\end{figure}

Indeed for the case $g=0$ {($h\neq 0$)} there is for any system size an immediate energy decrease in the hot subsystem, as shown in Fig. \ref{fig:IsingBumpheight} (top). This case is classical with only a small quantum tunneling between the two subsystems.
For generic values of $g$ and $h$, on the other hand, there is an initial rise in energy in accordance with the universal relation Eq.\eqref{eq:1}. The height of the energy bump $(E_m)$ is now independent of $N$, due to the {more local point-like interaction compared to the SYK non-local all-to-all tunneling.} 
This 
suggests that
the bump energy per particle $(E_m/N)$ will 
vanish in the thermodynamic limit to match our classical intuition. However, instead of such a thermodynamic vanishing, we should expect that also a finite size system exist where semi-classical hot-to-cold energy dynamics overwhelms the information-driven gain at short times.
Indeed for a fixed temperature, we can estimate where the bump disappears, by
extrapolating the $E_{m}/N$ to 
large $N$. Now we see the foretold disappearance of the bump at a fixed finite temperature at a finite value of $N$, restoring our classical intuition (Fig.\ref{fig:IsingBumpheight}).
An explicit finite $N$ example is given in Fig.\ref{fig:fixedptIsing}. 
This finite $N$ example shows that it is not simply the fact that the interaction is local and thus non-extensive in the thermodynamic limit, that causes it to vanish for higher temperatures.

The most interesting case is the conformal point of the Ising model (Fig. \textcolor{red}{\ref{fig:IsingBumpheight}}) (see also \cite{Kormos_2017,Perfetto_2017}). At exactly $g=1, h=0$ the bump only disappears by extrapolation to the continuum limit, similar to the SYK$_4$ results. This is still consistent with the earlier results on quenched cooling in conformal systems \cite{Bernard:2012je,Bernard:2013aru,Bhaseen:2013ypa}. The absence of a bump found there relies on conformal symmetry which is only a true symmetry in the continuum limit. 
At the same time for any finite size quantum system at low $T$, there appears to always be a small but non-zero counterintuitive initial rise. The bump is a correlation driven effect, as a simple ballistic collision model based on the Boltzmann equation will never have an initial energy rise in the hot system \cite{Doyon:2014qsa}.\footnote{Perhaps the easiest way to see this is to realize that the quenched cooling protocol is the quantum version of the Riemann problem in hydrodynamics. In hydrodynamics one assumes local equilibrium and thus an absence of correlations between different spatial points at distances larger than the local mean free path.} The correlation can still be either quantum or classical statistical. In the latter case, this classical statistical two-particle correlation (the two-particle distribution function) vanishes in the thermodynamic limit in accordance with the assumption of molecular chaos.

In summary, classical thermodynamics --- or rather hydrodynamics as we are studying time-dependent processes --- emerges in the quasi-particle (high temperature low density) limit with a non-extensive interaction between system and reservoir and after taking the thermodynamic limit.
The converse is that in quantum systems the 
initial rise 
in energy in the hot system that undergoes quenched cooling is robust and generic, though not required, and  universally explained by Eq.\eqref{eq:1a}.

\section{Conclusion}

In this manuscript we have analyzed the origins of the observed counter-intuitive early time energy increase in hotter systems {quench-coupled} to a cooler reservoir in quantum simulations. 
Our numerical study of Majorana SYK$_4$, using exact diagonalization demonstrates that the early time energy behaviour is proportional to the increase of the von Neumann entropy and is not related to a thermal flux from the cold to the hot system, {demonstrating} the quantum nature of this phenomenon. 
The energy increase is counterbalanced by the negative interaction potential (expectation value of the tunneling term in the Hamiltonian). In the setup here, the coupling quench does not supply energy into the system and the total energy is conserved. The same potential sets the amount of work needed to decouple the systems at given later time.

This peculiar phenomenon is well explained by the quantum {non-equilibrium extension} of the first law of thermodynamics Eq.\eqref{eq:1a} where the relative entropy $D(\rho(t)||\rho_T)$ plays a crucial role. Starting from a thermal state $D(\rho(t=0)||\rho_T)=0$ and using the positive semi-definiteness $D(\rho(t)||\rho_T)\geq 0$  the von Neumann entropy, scaled by the initial temperature, then sets a lower bound on the energy in each subsystem \eqref{eq:1}. This links the observed energy increase even in the hotter subsystem to an increase of the von Neumann entropy. Moreover, at sufficiently early times the change of the relative entropy is negligible compared to the energy which has two interesting consequences. Firstly, the early time evolution of the energy is almost directly proportional to the von Neumann entropy as we emphasized in our earlier paper \cite{gnezdilov2021information}; this provides a way to measure (dynamical) entanglement between two subsystems.\footnote{As the relative entropy is a measure of how distinguishable two states are, extremely small relative entropy means that at early times the subsystem is nearly indistinguishable from its initial thermal state implying that the energy increase is not related to a temperature rise, contrary to what was suggested in other papers {\cite{Zhang2019Evaporation,Almheiri2019Universal}}.}
{Secondly, it proves that the initial thermal state isn't instantaneously destroyed, hence the initial energy rise is not related to a temperature increase.}

The universality of this bound gives rise to an even more puzzling question: Why is such an energy increase not commonly encountered in our daily life? The reason lies in the quantum nature of this phenomenon. We show that at high temperatures in weakly interacting quasi-particle systems the height of the bump is suppressed and the time it crests  gets very short. In the thermodynamic limit it vanishes altogether, making it essentially {unnoticeable at everyday macroscopic scales.} 
As our results for SYK and the conformal point of the mixed field Ising model show, the more quantum mechanical the system is the closer one must push to the continuum quasiparticle limit for this bump to disappear and classical intuition to be restored. By extrapolation of our numerical simulation this is only ever possible to achieve in the strict thermodynamic limit.

This energy increase of the hotter system defies our intuition and understanding of classical thermodynamics but, as demonstrated here, it is well in accord with the laws of quantum thermodynamics.

There are three notable considerations that follow: There has been an substantial amount of research in the past few years on the out-of-time-ordered correlation function as a probe of classical and quantum chaos resulting in information exchange, scrambling and entropy growth (see e.g. \cite{2018NatPh..14..988S}). The standard wisdom is that this information flow is separate and faster than energy flow, because the latter is constrained by a conservation equation, as recalled for instance in \cite{Khemani:2017nda}. The result here and particular the inequality Eq.\eqref{eq:1} shows that this information flow, even though it is faster, must always drag some energy with it.

Secondly, one of the motivations to study SYK quenched cooling has been the equivalence with black hole evaporation through the holographic AdS/CFT correspondence. Because the evaporation of the black hole must expose the information behind the horizon, the quench can be modeled in the black hole context by a negative energy shock wave \cite{Engelsoy:2016xyb,Almheiri:2018xdw}, which shrinks the horizon upon contact. 
The result here shows that at very early times (before the shock hits the horizon in global time), there should be an interesting connection between the Ryu-Takayanagi entanglement surface encoding the von-Neumann entropy and the dynamics of the energy wavefront that holographically encodes Eq.\eqref{eq:1}.
                 
Finally, as already emphasized in \cite{gnezdilov2021information}, the inequality Eq.\eqref{eq:1} saturates in perturbation theory and can therefore be used in quenched cooling of weakly coupled systems to  probe the von-Neumann entropy.  Moreover, this is a universal result in the short time scale regime which is normally considered too sensitive to {peculiar details of the experimental set-up and the system} to be of interest. {It invites an experimental measurement of this universal way the von-Neumann entropy determines the energy response.}

\textbf{Acknowledgements} --- We thank Jan Zaanen for discussions in the early stage of this project and we thank Sebastian Deffner and Akram Touil for discussions.  This research was supported in part by the Netherlands Organization for Scientific Research/Ministry of Science and Education (NWO/OCW), by the European Research Council (ERC) under the European Union’s Horizon 2020 research and innovation programme.

\bibliography{SYKquenchrefs} 

\end{document}